\documentclass[aps,pra,preprint,showpacs,superscriptaddress,amsfonts,floatfix]{revtex4}
\usepackage{graphics}

\usepackage[dvips]{graphicx}
\usepackage{amsmath}

\begin{document}

%
%

\title{Two-channel approach to the average retarding force of metals for slow singly ionized projectiles}

%
%

%
\author{I. Nagy}
\affiliation{Department of Theoretical Physics,
Institute of Physics, \\
Budapest University of Technology and Economics, \\ H-1521 Budapest, Hungary}
\affiliation{Donostia International Physics Center, P. Manuel de
Lardizabal 4, \\ E-20018 San Sebasti\'an, Spain}
\author{ I. Aldazabal}
\affiliation{Centro de F\'{i}sica de Materiales (CSIC-UPV/EHU)-MPC,
P. Manuel de Lardizabal 5, \\ E-20018 San Sebasti\'an, Spain}
\affiliation{Donostia International Physics Center, P. Manuel de
Lardizabal 4, \\ E-20018 San Sebasti\'an, Spain}
%

%
\begin{abstract}

Based on the fundamental momentum-transfer theorem [Phys. Rev. Lett. {\bf 15}, 11 (1965)] 
a novel contribution to the retarding force of metallic systems for slow intruders is derived.
This contribution is associated with sudden charge-changing cycles during the path of projectiles.
The sum of the novel and the well-known conventional contributions, both expressed in terms of scattering phase shifts,  are used to discuss experimental data obtained for different targets. 
It is found that our two-channel modeling, with two nonlinear channels, improves the agreement
between several data and theory and thus, as predictive modeling, can contribute to the desired convergence between experimental and theoretical attempts on the retarding force.
 
\end{abstract}


\maketitle

\newpage

\section{Introduction and Motivation}

According to a basic book on quantum mechanics by Landau  \cite{Landau58} one of the most
important quantities in the interaction of {\it heavy} charged projectiles with fixed atoms is the average
energy loss. This time-independent quantity, a kind of deposited energy, is an observable and due to conservation laws its measurement is feasible in experiments. Thus in this subfield of nature (physics, human therapy) the real challenge resides in the {\it convergence} of measurements and theories.  Their interplay, a continuous one over a century,  fertilizes the developments
on both side of approaches which can  result in a transferable knowledge \cite{Sigmund14}.

The present contribution is dedicated to the case where singly ionized projectiles interact with constituents of metallic targets. The main challenge addressed here is to find a reasonable combination 
of the quantum statistical and atomistic aspects of the energy loss process in real targets.
As motivation, on which our new attempt is partially based, we start with an established result.
The well-known, conventional form \cite{Bonderup81,Zwicknagel99,Nagy04} for the stopping power
(written in Hartree atomic units, where  $e^2=m_e=\hbar=1$) of a homogeneous
degenerate electron gas (characterized by Fermi velocity $v_F$) for heavy intruders is given by 
\begin{equation}
\frac{dE}{dx}=\frac{2}{(2\pi)^3}\, \int_{0}^{v_F}\, v_e^2\, \left[2\pi\,  \int_{-1}^{1}dx\, (v-v_e\, x)\, v_r\, \sigma_{tr}(v_r)\right]\, dv_e,
\end{equation}
where $v$ and $v_e\in{[0,v_F]}$ are the projectile and system-electron velocities. 
One of the integration-variables is $x\equiv{\cos \beta}$, where $\beta$ is the angle between
$\bf v$ and ${\bf v}_e$. Thus $v_r^2=(v^2+v_e^2-2\, v_e v\, x)$.
Clearly, in an interpretation based on independent-electron scattering off a heavy projectile 
moving with constant velocity $v$, the remaining task resides in a {\it common} two-body interaction 
$V(r)$, in order to perform the statistical averaging over a Fermi-Dirac distribution with
\begin{equation}
\sigma_{tr}(v_r)=\frac{4\pi}{v_r^2}\, \sum_{l=0}^{\infty}(l+1)\sin^2[\delta_l(v_r)-\delta_{l+1}(v_r)].
\end{equation}

In this scattering interpretation the analysis is based on the concept of asymptotic states in the
infinite past and future, i.e., involving large time differences. Sudden processes in time, like a local charge-change in metals, requires a refined approach on associated transition amplitudes in time-dependent
perturbation theory. In kick-like sudden \cite{Landau58} processes one may use predetermined states 
as a complete set to treat the matrix elements in strong (but transient in time) perturbations. Notice that 
precisely it is such a transient channel which could make
difficulties in large-scale simulations, like in the orbital-based implementation of Time-Dependent Density-Functional Theory (TDDFT).  

Such an implementation rests on averaging of quantum mechanical time-dependent energy differences over certain time scales in order to  define a force-like quantity as stationary observable in stopping \cite{Zeb13,Correa15}. In these large-scale simulations specification of the initial conditions is required to real-time propagation. For instance, in \cite{Correa15} two options were considered for helium in aluminum target. In the first one, the screened atom was included in the determination of the static initial state. In the second one, the initial condition was set up by adding an $\alpha$-particle and thus producing a sudden change in the external potential. In both cases the authors control only the initial state and not the subsequent dynamics which is given by the time-dependent single-particle equations within TDDFT. Therefore, a smoothed evolving picture, without fast local charge-changing processes, is employed.

First, as concretization of motivation,
we integrate Eq.(1) by using models for the momentum transfer  cross section in order to get useful
informations to phenomenology made in Section II after Eqs.(6). Namely, we take the form of
$\sigma_{tr}(v_r)=4\pi A_{\alpha}/v_r^{\alpha}$, in which $\alpha=2$ and $\alpha=4$.
By straightforward quadrature we obtain \cite{Nagy04} from Eq.(1) for these models 
\begin{equation}
\frac{dE}{dx}=\frac{4\pi A_2}{(v_F)^2}\, n_0\, v\, v_F\left[1-\frac{1}{5}\left(\frac{v}{v_F}\right)^2\right] = A_2\, 
\frac{4}{3\pi} v_F^2\, \, v \,
\left[1-\frac{1}{5}\left(\frac{v}{v_F}\right)^2\right] \, \, \, for \, \, \, \, v\leq{v_F},
\nonumber
\end{equation}
\begin{equation}
\frac{dE}{dx}=\frac{4\pi A_2}{(v)^2}\, n_0\, v^2\left[1-\frac{1}{5}\left(\frac{v_F}{v}\right)^2\right] \, \, \, for \, \, \, \, v\geq{v_F}.
\nonumber
\end{equation}
\vskip 2mm
\begin{equation}
\frac{dE}{dx}=\frac{4\pi A_4}{(v_F)^4}\, n_0\, v\, v_F = A_4\, \frac{4}{3\pi}\, v
\, \, \, for \, \, \, \, v\leq{v_F}, \nonumber
\end{equation}
\vskip 0.5mm
\begin{equation}
\frac{dE}{dx}=\frac{4\pi A_4}{(v)^4}\, n_0\, v^2 \, \, \, for \, \, \, \, v\geq{v_F}. \nonumber
\end{equation}
The above model results are, of course, in agreement with expected limits $(dE/dx)=n_0vv_F\sigma_{tr}(v_F)$ and  $(dE/dx)=n_0v^2\sigma_{tr}(v)$, at $v\rightarrow{0}$ and $v\rightarrow{\infty}$, respectively.
Earlier, careful theoretical analysis \cite{Salin99} performed within an adiabatic framework on velocity-dependence states that the next term beyond the $v$-proportional one is {\it at least} second order in velocity. Our closed expressions for $v\leq{v_F}$ are in harmony with this important statement. Furthermore, a certain
weighted combination of our two expressions at $v\leq{v_F}$ would result in an almost perfect $v$-proportionality.
That, at this point simple mathematical, observation will become a more transparent and physical one in Section II,
where we extend the theory on the average retarding force beyond the common fixed-potential approximation by considering physically {\it reasonable} force matrix elements as independent channel contributions.

The rest of this paper is organized as follows. Section II is devoted to the theory and the discussion
of the results obtained. The last Section contains a short summary and few dedicated comments. As above, we
use atomic units throughout this work.


\section{Results and discussion}

We begin this section by outlining few important elements of stationary scattering theory. According to basic rules of quantum mechanics on expectation values of operators, one should consider the force matrix element \cite{Lippmann65} between orthonormal components of a scattering state to characterize  the associated momentum transfer. Applying this quantum mechanical theorem, where 
$\sigma_{tr}(v_r)\propto{\sum_{l=0}^{\infty}(l+1)[I_1(l,v_r)]^2}$,
one has \cite{Gaspari72,Bonig89,March19} for the matrix elements
\begin{equation}
I_1(l,v_r)=
\left[\int_0^{\infty}dr r^2 R_l(r,v_r)
\frac{\partial{V(r)}}{\partial{r}} R_{l+1}(r,v_r)\right]=
\sin[\delta_l(v_r)-\delta_{l+1}(v_r)].
\end{equation}
We stress that this remarkable {\it identity} rests on those states which refer to
the scattering Schr\"odinger wave equation with $v_r^2/2$ energy and $V(r)$ external field.
However, with partial waves based on $V(r)$, but with a net Coulomb field $\Delta V_c(r)=-q/r$
in space of $V(r)$ we get
\begin{equation}
I_2(l,v_r)=
\left[\int_0^{\infty}dr r^2 R_l(r,v_r)
\frac{\partial}{\partial r}\left(-\frac{q}{r}\right) R_{l+1}(r,v_r)\right]=
\frac{q\, \cos[\delta_l(v_r)-\delta_{l+1}(v_r)]}{2 v_r (l+1)}, 
\end{equation} 
and with unperturbed ($u$) partial wave components the corresponding result becomes
\begin{equation}
I_2^{(u)}(l,v_r)=
\left[\int_0^{\infty}dr r^2  j_l(v_r r)
\frac{\partial}{\partial r}\left(-\frac{q}{r}\right) j_{l+1}(v_r r)\right]=
\frac{q}{2 v_r (l+1)}. 
\end{equation}
Here we used the spherical Bessel functions of the first kind, i.e., the components of an unperturbed
plane-wave (momentum) state, instead of self-consistent radial functions. These forms in Eqs.(4-5) are based
on the fact that in cases with {\it abrupt} perturbations the original stationary system has no time \cite{Landau58} to relax to the stationary state of a new Hamiltonian.

We will consider these amplitudes as the proper ones when there is a sudden change in the 
self-consistent $V(r)$, as in the case of {\it charge-changing} ($q=1$) processes generated by the 
binary interaction with fixed lattice ions. This charge-change results in an excess bare field $\Delta V_c(r)=-1/r$.
The square of $[I_2(l,v_r)-I_2^{(u)}(l,v_r)]$
can characterize, in a quantum mechanical interpretation, 
an extra (kick-like) momentum transfer due to the sudden change in the external field.
That square is, in fact, a regularized transition probability. Such a regularization is needed 
since both $I_2$ and $I_2^{(u)}$ would give divergent results after $l$-summation. 
This regularized channel  gives (at $q\neq{0}$) a novel form for the associated cross section
\begin{equation}
\sigma_{tr}^{(2)}(v_r)=\frac{4\pi}{v_r^2}\,  \left(\frac{q}{v_r}\right)^2\,  \sum_{l=0}^{\infty}\frac{1}{l+1}\, 
\sin^4\left[\frac{\delta_l(v_r)-\delta_{l+1}(v_r)}{2}\right],
\end{equation}
to which a simple trigonometrical identity $(1-\cos \alpha)^2=4\sin^4(\alpha/2)$ is employed. 


Before our quantitative analysis, we continue with phenomenology. 
There are important differences between Eq.(6) and Eq.(2), i.e., between $\sigma_{tr}^{(2)}(v_r)$ and the conventional one given by Eq.(2) and denoted from here by 
$\sigma_{tr}^{(1)}(v_r)$. The kinematical 
prefactors show that 
the new term ($\propto{v_r^{-4}}$) vanishes faster at large scattering wave number ${v_r}$.
Thus  at $v_r\leq{1}$ values, which represent at small intruder
velocity the range of the Fermi velocity of metals, and at $\delta\simeq{\pi}$ for a dominating phase shift, the new term can become the dominating one. A combination of the $v_r^{-2}$ and $v_r^{-4}$
dependences in $[\sigma_{tr}^{(1)}(v_r)+\sigma_{tr}^{(2)}(v_r)]$
with the integrated characteristics found with separated model cross sections in the Introduction, signals that a
velocity-proportionality in the stopping power holds, practically up to $v\simeq{v_F}$ from below.


Now, we turn to the quantitative part of this Section. We will determine numerically the two quantities, denoted by 
$Q^{(1)}(v_F)$ and $Q^{(2)}(v_F)$, by which the low-velocity stopping power of metals (a system of an electron gas and lattice ions) takes a friction-like form
\begin{equation}
\frac{1}{v}\, \frac{dE}{dx}\, =\, \left[Q^{(1)}(v_F)\, + Q^{(2)}(v_F)\right],
\end{equation}
where the two coefficients (when $q\neq{0}$) are given by the following expressions
\begin{equation}
Q^{(1)}(v_F)\, =\, \frac{4}{3\pi}\, v_F^2\, \sum_{l=0}^{\infty}(l+1)\sin^2[\delta_l(v_F)-\delta_{l+1}(v_F)],
\end{equation}
\begin{equation}
Q^{(2)}(v_F)\, =\, \frac{4}{3\pi}\,  q^2\, \sum_{l=0}^{\infty}\frac{1}{l+1}\, 
\sin^4\left[\frac{\delta_l(v_F)-\delta_{l+1}(v_F)}{2}\right].
\end{equation}
Our summation of two channel cross sections in Eq.(7) resembles, mathematically, to the well-known \cite{Burke77} rule in potential scattering with a spin-orbit interaction term where we sum the direct (non-spin-flip) and spin-flip
partial differential cross sections for electron scattering for any spin orientation before scattering. There, the integrated
cross sections, needed to observables, are obtained by integrating over all scattering angles.
Remarkably, the spin-flip part depends on an  amplitude difference, similarly to our regularized difference. 

We stress at this point that we employ to summation in Eq.(7) an {\it a priori} unit-weight assumption. In reality,  i.e., at channeling-like conditions in metals, the impact parameter-dependent closest approach of intruders and lattice ions \cite{Correa15,Bergsmann98,Primetzhofer11,Tran19}
may influence that assumption. In more simple terms, our present weighting would refer to random-collision situations. Nonequal weighting might be based on certain probabilistic inputs \cite{Andres90} to sum {\it two} nonlinear channel. But, such inputs need, in our modeling, an additional justification, since one can not apply stationary linear-response ideas to a sudden effect.


%
\begin{table}
\caption{\label{tab:parameters} Partial contributions,  $Q^{(1)}(v_F)$ and   $Q^{(2)}(v_F) $ at $q=1$,
to Eq.(7).
Phase shifts, based on the orbital version of DFT \cite{Puska83,Echenique85,Nagy89}, are employed. See the text for further details.}
\begin{ruledtabular}
\begin{tabular}{ccccccc}
\ & \multicolumn{2}{c}{$r_s=1.5$ \qquad\ } &
\multicolumn{2}{c}{$r_s=2$ \qquad\ } &
\multicolumn{2}{c} {$r_s=3$}       \\
$Z_1$ & $Q^{(1)}$ & $Q^{(2)}$ & $Q^{(1)}$ & $Q^{(2)}$ & $Q^{(1)}$ & $Q^{(2)}$ \\
\hline
$1$ &  $0.305$ & $0$ &  $0.255$ & $0$ & $0.162$  & $0$ \\
$2$ &  $0.755$  & $0.069$ &  $0.427$ & $0.134$ & $0.135$ & $0.250$  \\
$3$ &  $0.912$ & $0.161$ &  $0.439$ & $0.247$ & $0.117$ & $0.368$   \\
$4$  &  $1.112$ & $0.235$ & $0.557$ & $0.323$ & $0.191$ & $0.421$   \\
$5$ &  $1.417$ & $0.298$ & $0.749$ & $0.374$ & $0.307$  & $0.443$    \\
$6$  &  $1.692$ & $0.366$ & $0.874$ & $0.413$ & $0.346$ & $0.481$   \\
$7$  &  $1.777$ & $0.369$ & $0.825$ & $0.449$ & $0.275$ & $0.522$   \\
$8$  &  $1.631$ & $0.402$ & $0.637$ & $0.483$ & $0.167$ & $0.545$   \\
$9$  &  $1.346$ & $0.438$ & $0.428$ & $0.512$ & $0.085$ & $0.563$   \\
$10$ &  $1.047$ & $0.471$ & $0.267$ & $0.539$ & $0.035$ & $0.593$   \\
$11$ &  $0.815$ & $0.498$ & $0.183$ & $0.564$ & $0.032$ & $0.612$   \\
$12$ &  $0.690$ & $0.520$ & $0.199$ & $0.572$ & $0.108$ & $0.595$   \\
$13$ &  $0.697$ & $0.531$ & $0.338$ & $0.559$ & $0.242$ & $0.546$   \\
$14$ &  $0.850$ & $0.533$ & $0.580$ & $0.527$ & $0.348$ & $0.484$   \\
$15$  & $1.146$ & $0.521$ & $0.846$ & $0.482$ & $0.360$ & $0.427$   \\   
$16$ & $1.539$ & $0.502$ & $1.062$ & $0.437$ & $0.297$ & $0.405$   \\   
$17$ & $1.975$ & $0.480$ & $1.219$ & $0.404$ & $0.234$ & $0.374$   \\   
$18$ & $2.386$ & $0.458$ & $1.364$ & $0.386$ & $0.191$ & $0.372$   \\   
\end{tabular}
\end{ruledtabular}
\end{table}

Table I contains our numerical results for  $Q^{(1)}(v_F)$ and $Q^{(2)}(v_F)$ at selected values of the
$r_s$ Wigner-Seitz radius and at $q=1$. The partial phase shifts, calculated by DFT at the Fermi momentum, are taken from earlier works \cite{Puska83,Echenique85,Nagy89}. Both $Q^{(1)}$ and 
$Q^{(2)}$ are oscillating functions. But $Q^{(1)}(r_s)$ has a strong, 
direct density-dependence via $v_F^2\propto{r_s^{-2}}$ in Eq.(8). 
Thus, at metallic densities,  $Q^{(2)}(r_s)$ in the sum $[Q^{(1)}+Q^{(2)}]$ can make an important modulation in the $Z_1$-oscillation of $Q^{(1)}$, especially around its minima.
For  $Z_1=1$, we take our values for $Q^{(1)}$ obtained within the explicit version \cite{Nagy04} of DFT. There a single Euler equation is solved in an iterative self-consistent  way. That calculation does not consider a doubly-populated weakly bound (extended) state around an embedded proton in an electron gas, in harmony with experimental facts, obtained by positive muons, on the nonexistence of muonium in metals. 

Despite this, there is a perfect agreement \cite{Nagy04} with $Q^{(1)}$ results obtained from the implicit, orbital-based,  DFT. This agreement signals that it is the short-range part of proton-screening which needs a nonlinear  treatment. In simple terms, that range is  the most important one to determine the first few phase shifts. Our $Q^{(2)}=0$ values for $Z_1=1$ are in accord with a screened-proton picture without bound state, where there is no charge-changing cycle, thus $q=0$ during the motion of the projectile. Since the experimental data, obtained with low velocity proton
projectiles for Al and Ni targets, are in very reasonable 
harmony \cite{Primetzhofer11,Tran19,Sortica19} with nonlinear theory \cite{Puska83,Echenique85,Nagy89} based solely on $Q^{(1)}(v_F)$, we have a transferable knowledge in this case. A desired {\it convergence} between the two sides of understanding is achieved.

For all other $Z_1\geq{2}$ we consider, for velocities $v\leq{v_F}$, the $q=1$ value as the most plausible one. This conservative value seems to be a realistic one with singly ionized intruders. 
Higher $q$
values might have relevance when there is a large electronic overlap between clouds of colliding atoms. We believe that such a partial channel with $q>1$ would need more energetic, head-on-like collisions. Theoretically, it would be interesting to model the 
{\it transition} from our discrete-$q$ modeling of charge-changing cycles to the pioneering \cite{Firsov59}
quasiclassical work where the retarding force (the observable) is related to an electron-density-flux constructed from the statistical Thomas-Fermi theory of atoms. 
With a transition-study one might arrive at a deeper understanding of an integrated (classical trajectory Monte Carlo) approach \cite{Gruner04} for energy loss and capture processes.

There is certain contradiction (c.f., Fig. 2, below), for Al target between low-velocity experiment \cite{Primetzhofer11} and TDDFT \cite{Zeb13,Correa15,Maliyov20} results in the case of  $He^{+}$ projectiles. In this case our novel result, based on [$Q^{(1)}+Q^{(2)}$], is in harmony  with \cite{Correa15} for the off-channeling situation. For the channeling simulation our $Q^{(1)}$ also gives
a very reasonable agreement with \cite{Zeb13,Correa15}. 
We stress that \cite{Maliyov20} uses
$\alpha$-particle ($He^{++}$) as projectile and an atom-centered optimized Gaussian basis set to model the energy transfer. The observed agreements (see, the discussion at Fig. 2) are especially remarkable in the light of careful experiment \cite{Riccardi15} performed on electron emission from aluminum. There a perfect linearity in the velocity of helium ions, with $v\leq{0.6}$, was obtained, and thus a quantitative agreement with \cite{Zeb13} was concluded. 

Notice, in the spirit of discussion made already in \cite{Correa15} for proton and helium intruders, that we can image experimental situation where, at very low ion velocities, only the neutral screened-atom gives contribution via its $Q^{(1)}$, and the $Q^{(2)}$ channel
becomes active only from an intermediate velocity below the Fermi velocity. Such a modeling would fit to the experimental \cite{Primetzhofer11} suggestion on two, both linear in ion-velocity, parts on the whole kinematical range $v\leq{v_F}$. Of course, the acceptance of such a suggestion presupposes that the
underlying experiment-evaluation method behind data is a well-justified one for the whole range of ion velocity. For these simple intruders, a
partial support to such a view could be based on surface-scattering experiment \cite{Winter03} performed with singly ionized ions ($Z_1\in{[1,20]}$ and at projectile velocity $v=0.5$) scattered from an aluminum surface at variable scattering angles. There, the challenging problem  of inhomogeneity in the electron density profile of the electron salvage in front of a metal surface was studied, with an accompanied phenomenological refinement of factors in $(dE/dx)=(n_0 v)[v_F\sigma_{tr}^{(1)}(v_F)]$.
We will return to this experiment, at the discussion of Fig. 3 which is devoted to an important  comparison for $Z_1>2$.

Related to our prediction with $[Q^{(1)}+Q^{(2)}]>Q^{(1)}$ values for the average retarding force in metals, we turn to a brief discussion of data \cite{Bergsmann98} obtained for an other free-electron-like material, Mg. For this metal $v_F\simeq{0.7}$ and the experiment with $He^+$  and proton projectiles was performed for $v\geq{v_F}$. It was found that the {\it ratio} ($R$) of stopping powers with these intruders
becomes about two, in contrast to a ratio of about unity which is based on $Q^{(1)}$ values of self-consistent DFT. Our novel approach would result in a ratio ($R>{2}$)  which is not in contradiction  with experimental suggestion. As support, we note that in \cite{Winter03}, i.e., in surface experiment, the helium per proton stopping ratio was found to be {\it always} higher than unity, even for $r_s(z)>3$.
Clearly, the desired {\it convergence} of theories and experiments requires further studies for 
Mg  ($r_s\simeq{2.7}$) and, say, for Ca ($r_s\simeq{3}$) as well, within large-scale TDDFT simulations with proton and helium intruders at $v\leq{v_F}$.


At this point, i.e., before the presentation and discussion of our illustrative Figures, we would like to
mention a very recent attempt \cite{Gomez20} where a {\it two-channel} modeling was presented for the spectral line-width in plasma environments. The authors of that insightful work demonstrated that the commonly used expression for the line-width neglects a potentially important contribution from electron-capture processes. Their numerical value signals that a proper sum of two contributions can be about
twice of the conventional estimation. In the field of high-energy-density plasmas, our $q$-mediated
enhancement in stopping power may contribute to the proper determination of the ignition threshold
\cite{Frenje19} in a deuterium-tritium-alpha energy deposition process. There, via a plausible postulation, the theoretical {\it underprediction} of stopping data was associated \cite{Frenje19} with ion-ion nuclear scattering.

Now, we illustrate our novel results by three Figures. In Fig.1, for $r_s=1.5$ of the Wigner-Seitz parameter, we plot the dimensionless ratios of $R_1=[Q^{(1)}(v_F,Z_1)/Q^{(1)}(v_F,Z_1=1)]^{1/2}$ and 
$R_2=\{[Q^{(1)}(v_F,Z_1)+Q^{(2)}(v_F,Z_1)]/Q^{(1)}(v_F,Z_1=1)\}^{1/2}$, i.e., ratios of nonlinear quantities. One might consider \cite{Sortica19,Matias19} these ratios as a kind of effective charge. This Figure reflects, in a highly phenomenological manner, that the so-called $Z_1$ oscillations may get important modulations especially around the minima of the conventional $R_1$ ratio.
Notice that the such-defined ratios are square roots of physical magnitudes. This mathematical
operation has a smoothing character (c.f., Fig. 3) with renormalized oscillating functions.

\begin{figure}
\scalebox{0.4}[0.4] {\includegraphics{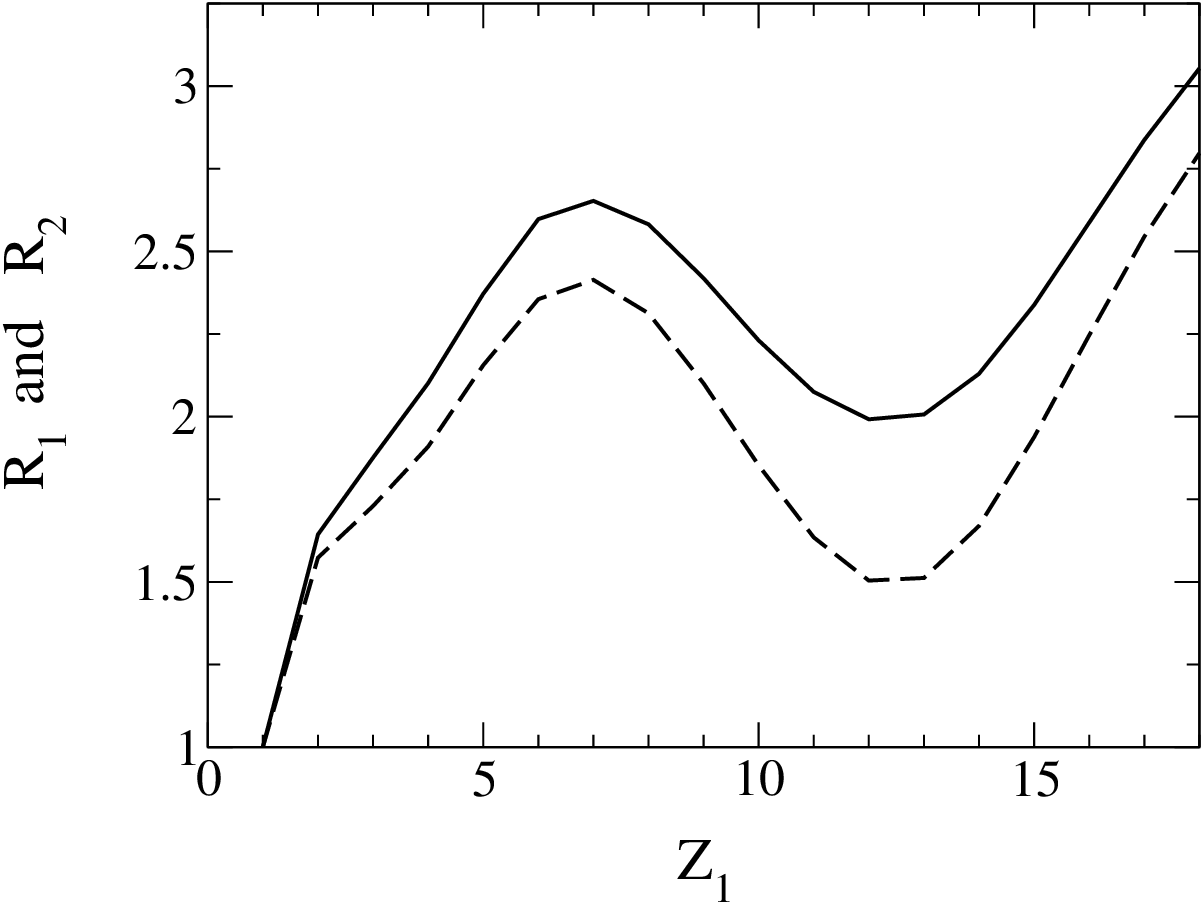}}
\caption{Illustrative dimensionless ratios, $R_1$ (dashed curve) and $R_2$ (solid curve), defined in the text, as a function of $Z_1$. The density parameter is $r_s=1.5$, which corresponds to a high-density degenerate electron gas. This would refer to the plasma frequency of  Au.}
\label{figure1}. 
\end{figure}

\newpage

Fig. 2 is devoted to $(dE/dx)$ quantities, in atomic units, obtained for Al 
($r_s\simeq{2.13}$) with helium projectiles. The velocity-range, in atomic units, is $v\in{[0,0.6]}$. 
The green dotted curve refers to a simple (with $Z_1=2$) linear-response (first-order Born) approximation, where the electron gas dielectric function at the RPA level, i.e., without static or dynamic local-field corrections \cite{Nagy85}, is used.
In such an approximate theory, the stopping is proportional to $Z_1^2$.  The corresponding form, employed in a cornerstone work \cite{Echenique85} as well, is given by
\begin{equation}
\left(\frac{dE}{dx}\right)_{RPA}=v\, \frac{4 }{3\pi}\, \left\{\frac{Z_1^2}{2}
\left[\ln \left(a+\frac{2}{3}\right)- \frac{3a-1}{3a+2}\right] 
\left(\frac{3a}{3a-1}\right)^2\right\},
\end{equation}
where $a=\pi v_F$ is an abbreviation. Its value is about $a\simeq{2.83}$ for Al. This expression is a particular
realization of the modeling made in the Introduction with $\sigma_{tr}(v_r)\propto{(1/v_r^{4})}$.

Our present results (at $Z_1=2$) correspond to the green solid curve $[Q^{(1)}+Q^{(2)}]$, and the green dashed curve [$Q^{(1)}$]. 
The black solid and red dash-dotted curves with filled circles
are taken from Fig. 5b of \cite{Correa15}. 
They refer to off-channeling and channeling conditions, respectively.
Notice that an earlier
TDDFT result of \cite{Zeb13} (not shown here) agrees precisely with the black curve. 
There is a fortuitous {\it similarity} between the RPA result for the homogeneous electron gas and results
plotted via curves by solid green and dash-dotted red with dots. Neither the screening-treatment nor the
scattering-description of RPA is correct to a nonlinear situation. For proton, where a very reasonable
agreement \cite{Primetzhofer11,Tran19,Sortica19} between nonlinear $Q^{(1)}(v_F)$ and data was found, the above Eq.(10) with $Z_1=1$ would  give a serious {\it underestimation} \cite{Echenique85}.

Experimental \cite{Primetzhofer11} data, used already in TDDFT to comparison \cite{Correa15}, are plotted here by a dashed  magenta curve with filled triangles. This curve signals a two-slope behavior with linearities in projectile velocity. Remarkably, a quite similar, i.e., two-slope, behavior was found in \cite{Tran19} for Ni ($r_s\simeq{1.8}$) with singly-ionized $He^{+}$ intruder. There, a comparison with TDDFT results \cite{Correa18} is made, by
using $1.15$ as multiplying factor for the simulation-based results.
As we already discussed above, we can image such a two-slope behavior within the present theoretical framework with certain, presumably closest-approach-dependent \cite{Nagy12}, finer tuning of our two nonlinear channels. A complete {\it convergence} is still not achieved.
The two black squares, for $\alpha$-projectile, are taken from Fig. 5b of \cite{Maliyov20}, for our velocity range. They are based on TDDFT with an optimized, localized atomistic, Gaussian basis set. We speculate that, for extended systems with slow ions, the screening action of the metallic electron gas needs a further consideration. Moreover,
singly ionized $He^{+}$ intruder, instead of $He^{++}$,  might be more close to the experimental situation.

\begin{figure}
\scalebox{0.4}[0.4] {\includegraphics{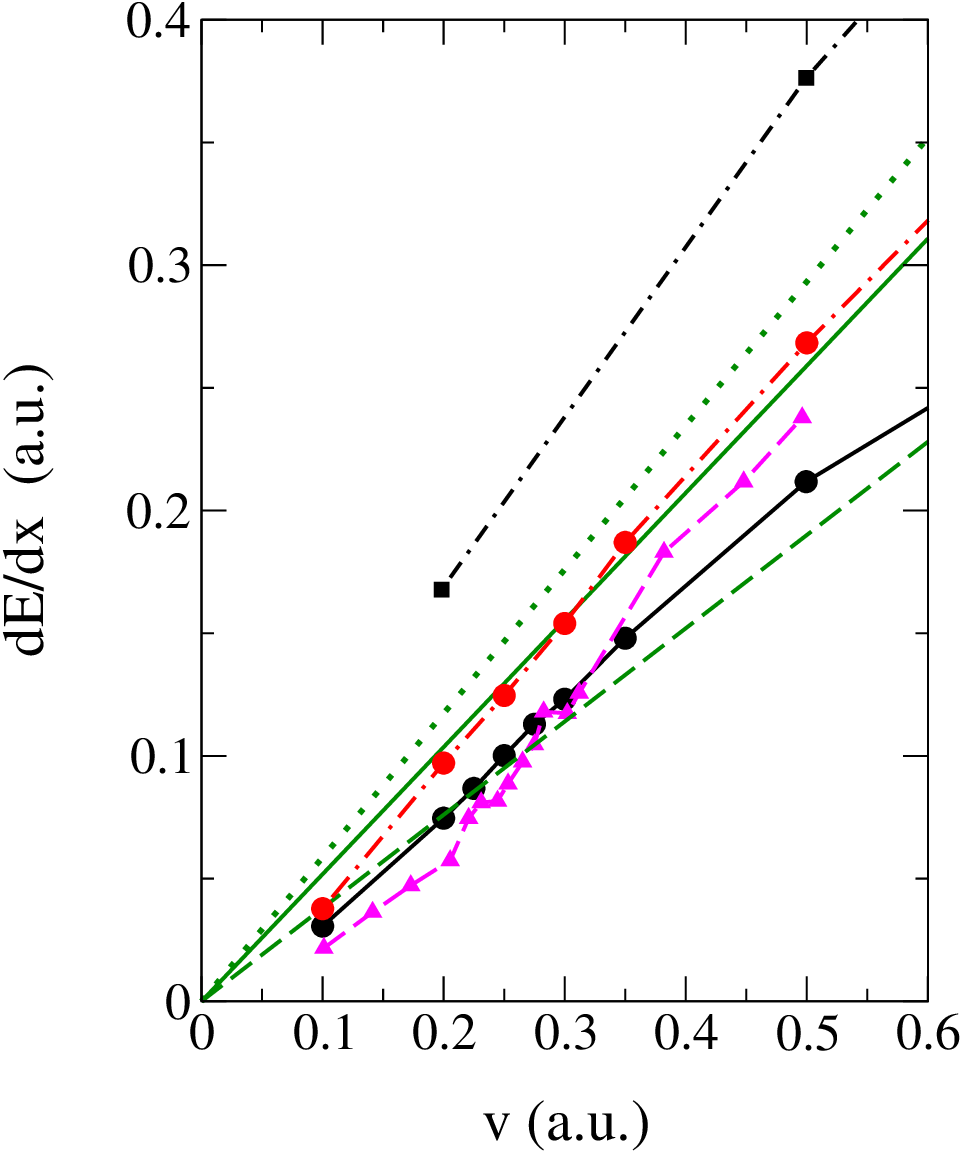}}
\caption{(Color online) Stopping power, $dE/dx$, for helium projectiles as a function of the velocity $v\in{[0,0.6]}$. The present results refer to  the green solid and green dashed curves.
Symbols are taken from Fig. 5b of \cite{Correa15} (black solid and red dash-dotted curves with filled circles), and from Fig. 5b of \cite{Maliyov20} (black squares).
Data for Al are plotted by a dashed  magenta curve with filled triangles.  Their systematic and statistical errors are analyzed in the experimental work \cite{Primetzhofer11}. Finally, the green dotted curve refers to Eq.(10) with $Z_1=2$. See the text for further details.} 
\label{figure2}. 
\end{figure}
%


In Fig. 3 we plot the observable quantities, $(1/v)(dE/dx)$, as a function of $Z_1$. The experimental data (black dots and triangles) are used earlier \cite{Echenique85} to a comparison
with $Q^{(1)}$, which is denoted here by a dashed curve. It was stated in this pioneering work that there is a substantial
disagreement with data in magnitude, particularly around the minimum. Our new result,
$[Q^{(1)}+Q^{(2)}]$, is denoted by a solid curve. Notice that data symbols, without error bars, refer to
$v=0.411$ (dots) and $v=0.826$ (triangles). The target is the frequently used prototype of free-electron 
metal, aluminum. By inspection,
one can observe an essential improvement in agreement between data and the novel approach. 
Here we return to experiment in \cite{Winter03}, i.e., to the above-mentioned surface experiment. There, although with somewhat smaller deviations from the conventional  $Q^{(1)}[v_F(z)]$-type scaling, 
also a systematic {\it upward} enhancement in stopping power was established. In the present 
two-channel modeling,
such an enhancement  can be associated with a $Q^{(2)}$-proportional contribution.

\begin{figure} 
\scalebox{0.4}[0.4] {\includegraphics{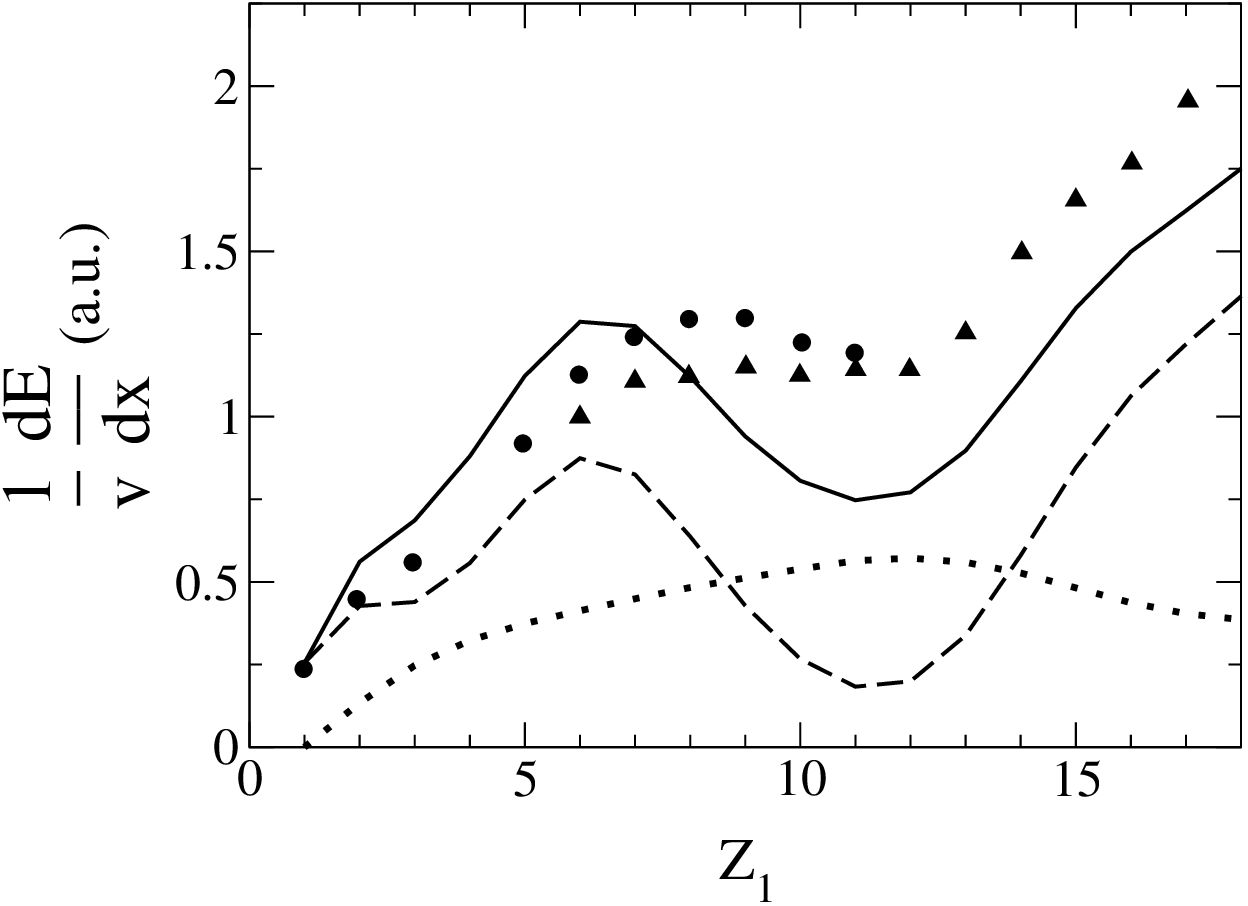}}
\caption{Results for the friction-like coefficients $(1/v)(dE/dx)$
are plotted in atomic units. 
The dashed and dotted curves refer to Eq. (8) and Eq. (9), respectively.
The sum of Eq.(8) and Eq.(9), the solid curve, refers to our novel approach. The experimental data points (for Al) are the same which were used in a pioneering work leaded by
Echenique \cite{Echenique85}. They are denoted by solid circles and triangles.}
\label{figure3}. 
\end{figure}
%


\section{summary and comments}

In this theoretical paper we have investigated the problem of average retarding force of metallic targets for slow projectiles. Beyond the well-known contribution,  established for electron-slow-intruder
scattering in a degenerate electron gas, a novel contribution is derived which is associated with
charge-change cycles due to lattice ions. Our main closed result is given by Eq.(6), which is, in the terminology
of this sub-field of physics, a nonlinear form, similarly to the more conventional one in Eq.(2). These
forms are implemented here by standard phase shifts obtained by applying the orbital-based DFT to screening in an electron gas. For helium projectiles, we made comparisons with selected results of experiments \cite{Primetzhofer11} and large-scale simulations \cite{Zeb13,Correa15} in TDDFT for 
a free-electron-like metal, Al. With our novel contribution to the retarding force, the agreement with these  results is improved.

As we discussed in Section I, our novel channel describes a sudden perturbation which is not explicit in recent
TDDFT simulations. There a smoothed evolving picture is employed, without fast local charge-changing
processes. As Fig. 2 signals our green curves bracket the TDDFT outputs obtained for channeling and
off-channeling conditions within  an evolving picture. We believe that further efforts in TDDFT are
needed to tackle explicitly charge-changing processes. However, by construction, TDDFT simulations
are able to model the lattice-related details of realistic targets. In order to get further informations on the
capabilities of different theoretical methods the problem of projectiles in alloys could be an important one. There, different lattice ions could influence (presumably, due to different closest approaches) the charge-changing fast processes giving
an opportunity to see the advantage of our modeling over those based on a smoothed picture in the time domain.


A challenging problem in ratio-data-interpretation \cite{Bergsmann98} for Mg is discussed as well. Our two-channel-based result is in reasonable agreement with  data at around $v=v_F\simeq{0.7}$. The conventional theoretical estimation is not in agreement with experiment.
A detailed comparison with data for $Z_1\in{[1,18}]$ in Al is made, and improved agreement
is found. Based on these agreements, we suggest further efforts within TDDFT along these lines. 
The percentage differences (about $400\%$  and $40\%$, roughly) of the conventional $Q^{(1)}$ (dashed curve) and the new $[Q^{(1)}$+$Q^{(2)}]$ (solid curve) results in Fig. 3  for $Z_1=12$  in comparison with experimental data on pure Al target heralds that alloy-targets (see, above) could be relevant
candidates to understand differences between theoretical attempts discussed in this work.

Notice that a recent adiabatic modeling \cite{Matias19}, motivated by experiment in \cite{Sortica19},
also results in remarkable deviations from a simple modeling with $Q^{(1)}(v_F)$. There the density inhomogeneity was considered, via lattice-atom-volume averaging  of $Q^{(1)}[v_F(r)]$, 
as a modulating effect. 
Such an averaging was applied successfully  \cite{Nagy02} for stopping of swift $Z_1=\pm{1}$ in order to discuss the charge-sign effect in Si. In the theoretical modeling \cite{Matias19} a remarkable
similarity with experimental effective charges (defined at Fig. 1) was obtained in such a way. Furthermore, it was suggested that calculations within large-scale TDDFT simulations would be useful to demonstrate the strength of the underlying \cite{Matias19} approach. We share this suggestion for realistic TDDFT. The suggestion made above on an other important challenge with Mg ($r_s\simeq{2.7}$), or with Ca ($r_s\simeq{3}$) is in accord with this.

Thus, at this moment, we have two, i.e., $q$-dependent and inhomogeneity-dependent, effects
which result in enhancement in the electronic stopping power beyond the conventional, i.e., $Q^{(1)}$-dependent, theoretical estimation. Both seem to be, {\it a priori}, relevant in reality. Their proper
weights and interplay need future investigations. Cases with self-irradiated 
condition \cite{Lim16,Artacho18}
could be especially important in this ($q\neq{0}$) respect. For instance, Ni ions in Ni target \cite{Artacho18}. In such a {\it symmetric} case we can image (for a metal) even $q=2$ to our $Q^{(2)}$ channel. For $r_s\simeq{2}$, Ni ion with its $Z_1=28$ represents the second minimum in $Z_1$-oscillation \cite{Puska83,Nagy89}. 
In our modeling we get $[Q^{(1)}+Q^{(2)}]\simeq{[0.28+(q^2\times{ 0.72})]}$.  

At $q=2$ and $v=1$, one arrives at $[(Q^{(1)}+Q^{(2)})/Q^{(1)}]\simeq{11}$,
thus the corresponding stopping power would change steeply to about $(dE/dx)\simeq{160\, eV/A}$.
For transition metals, that show a high electronic stopping power \cite{Artacho18}, the spin-flip
process needs a thorough investigation. The electron spin direction is no longer conserved during
electron-atom collision. One has to consider the  total angular momentum ${\bf j}=({\bf l}+{\bf s})$ 
operator in order to construct a complete set of spin-angle functions which are needed to expansions. We left this exciting sub-problem in stopping theory with a new (spin) degree of freedom to future studies.

Notice that at high ion velocity, our new term would scale as $(q/v)^2$ with respect to the conventional, i.e., Bethe-like \cite{Landau58}, leading one \cite{Correa15,Maliyov20}. 
There, a term with $[q(v)/v]^2$ can give a {\it slowly} vanishing  enhancement.
Thus the  high-velocity limit, first of all  under self-irradiated condition \cite{Artacho18}, also requires further investigations. The precise relevance of permutation-based, similarity-aided
level crossing \cite{Firsov59,Migdal77} behind higher 
$q(v)$-values seems to be an other interesting sub-problem in stopping theory. The Bethe-limit, especially for metals with their dense electron gas, is not a simple cumulative sum of isolated atomic contributions \cite{Nobel05}.


Based on the established capability of our new modeling for metals, we believe that the two-channel approach developed here can find application in other important fields as well. For instance, in the friction problem of diatomic molecules during their dissociative adsorption on metallic surfaces. There, based on an empirically motivated local-density-friction approximation, a
local $Q^{(1)}[v_F(r)]$ is employed \cite{Juaristi17}. We argue here that transient electronic processes, due to dissociation in an electron gas, could be related to $Q^{(2)}$ in Eq.(6). For instance, the case of
N, with its $Z_1=11$, might be a good candidate, as our Table I suggests. We stress, however, that at high target-temperatures, the coupling to phonon modes, i.e., to quanta of lattice vibrations, can open a new  \cite{Gondre12} channel to inelastic processes, beyond the friction-like channel discussed in our present study for cold metals. Still,  as  Figure 1 of \cite{Gondre12} signals, the proper {\it magnitude} of this latter channel could be important. Indeed, the so-far neglected \cite{Gondre12} charge-transfer-type [related to $Q^{(2)}(v_F)$] processes, especially with highly reactive molecules, may have impact on conclusions.

We close with few general comments. The wave functions of the conventional orbital-based DFT for embedded $Z_1$ are used \cite{Puska83,Echenique85,Nagy89} here to calculate the induced electron density. That is the basic variable of the underlying variational theory. The phase shifts are, therefore, auxiliary quantities \cite{Puska83}. Their sums over angular momentum quantum numbers
always satisfy the associated neutrality condition of a self-consistent  orbital-based approximation, i.e., the Friedel sum rule and the Levinson theorem \cite{Burke77} for {\it local} interactions. Since these
are satisfied by construction for any form of a local many-body term in the Schr\"odinger-like equations, the physical quality of DFT results needs further, i.e., energetic, justifications.


But, in accord with closely related statements \cite{Puska83,Nagy89}, the highly improved quantitative agreements with experimental facts justify, {\it a posteriori}, our phase-shift-based two-channel modeling
with a novel term for the retarding force. Generally, and in accord with Landau basic attempt \cite{Migdal77} for Fermi liquids, 
a modeling is good if it contains few adjustable elements, agrees with several observations, and makes controllable predictions. We stress, finally,  that the truly exciting  theoretical problem of interparticle-interaction, i.e., correlated motion of electrons, is considered in stopping calculation only at the mean-field level. However, at least for a prototype two-particle correlated model system, recent exact result \cite{Nagy19} for the energy shift in time-dependent (passing)  perturbations indicates that proper independent  {\it modes}, rather than {\it effective} single-particle states, could pave the path to future developments.
 

%
\begin{acknowledgments}
One of us (IN) is indebted to P. M. Echenique for his continuous interest and advices.
We are thankful to P. Bauer, R. D\'iez Muino,  J. I. Juaristi, I. Maliyov, D. Primetzhofer, and
D. S\'anchez-Portal, for very useful discussions. This work was supported partly by the Spanish Ministry of Economy and Competitiveness (MINECO: Project FIS2016-76617-P).
\end{acknowledgments}
%


\end{document}